\documentclass[sn-mathphys,Numbered]{sn-jnl}

\usepackage{graphicx}%
\usepackage{multirow}%
\usepackage{amsmath,amssymb,amsfonts}%
\usepackage{amsthm}%
\usepackage{mathrsfs}%
\usepackage[title]{appendix}%
\usepackage{xcolor}%
\usepackage{textcomp}%
\usepackage{manyfoot}%
\usepackage{booktabs}%
\usepackage{algorithm}%
\usepackage{algorithmicx}%
\usepackage{algpseudocode}%
\usepackage{listings}%

\theoremstyle{thmstyleone}%
\theoremstyle{thmstyletwo}%

\theoremstyle{thmstylethree}%

\begin{document}

\title[Article Title]{Non-volatile Tuning of Cryogenic Optical Resonators}


\author[1]{\fnm{Uthkarsh} \sur{Adya}}\email{uthkarsh@uw.edu}

\author[1]{\fnm{Rui} \sur{Chen}}\email{charey@uw.edu}

\author[1]{\fnm{I-Tung} \sur{Chen}}\email{itungc@uw.edu}

\author[1]{\fnm{Sanskriti} \sur{Joshi}}\email{sjoshi3@uw.edu}

\author[1,2]{\fnm{Arka} \sur{Majumdar}}\email{arka@uw.edu}

\author[1,2]{\fnm{Mo} \sur{Li}}\email{moli96@uw.edu}

\author*[1]{\fnm{Sajjad} \sur{Moazeni}}\email{smoazeni@uw.edu}

\affil[1]{\orgdiv{Department of Electrical and Computer Engineering}, \orgname{University of Washington}, 
\orgaddress{\street{185 E Stevens Way}, \city{Seattle}, \postcode{98195}, \state{WA}, \country{USA}}}

\affil[2]{\orgdiv{Department of Physics}, \orgname{University of Washington}, 
\orgaddress{\street{185 E Stevens Way}, \city{Seattle}, \postcode{98195}, \state{WA}, \country{USA}}}


\abstract{Quantum computing, ultra-low-noise sensing, and high-energy physics experiments often rely on superconducting circuits or semiconductor qubits and devices operating at deep cryogenic temperatures (4K and below). Photonic integrated circuits and interconnects have been demonstrated for scalable communications and optical domain transduction in these systems. Due to energy and area constraints, many of these devices need enhanced light-matter interaction, provided by photonic resonators. A key challenge, however, for using these resonators is the sensitivity of resonance wavelength to process variations and thermal fluctuations. While thermo-optical tuning methods are typically employed at room temperature to mitigate this problem, the thermo-optic effect is ineffective at 4K. To address this issue, we demonstrate a non-volatile approach to tune the resonance of photonic resonators using integrated phase-change materials (PCMs) at cryogenic temperatures. In this work, we report a 10~Gb/s free-carrier dispersion based resonant photonic modulator that can be tuned in a non-volatile fashion at sub-4K temperatures using a commercial silicon photonics process. This method paves the way for realizing scalable cryogenic integrated photonics with thousands of resonant devices for quantum and high-energy physics applications.}





\maketitle

\section{Introduction}
\label{sec1}

Future cryogenic quantum and classical computing as well as high-energy physics (HEP) detector systems require optical interconnects to communicate data and control signals between room temperature (RT) and cryogenic stages~\cite{lecocq2021control, DUNE:2020ypp, PhysRevLett.125.260502}. Recent demonstrations have shown promising advantages of optical fiber links over electrical cables, including ultra-high bandwidths resulting in the ability to achieve terabits-per-second (Tb/s) data rates over long distances, minimal heat load transferred to the cryogenic stage, and negligible thermal noise~\cite{lecocq2021control,reilly2019challenges,youssefi2021cryogenic}. The key components enabling such optical links for practical and scalable systems are cryogenic photonic resonators and resonant modulators, which offer low-voltage and energy-efficient modulation within a compact footprint~\cite{Pintus:22,Gehl:17}. These resonators have also been directly interfaced with superconducting qubits and circuits for electro-optical transduction~\cite{mirhosseini2020superconducting,weaver2024integrated}. 

These resonators can be fabricated using advanced silicon photonics processes provided by commercial foundries, enabling the development of large-scale, practical photonic integrated circuits (PICs)~\cite{rakowski202045nm,siew2021review}. However, the major challenge in using photonic resonators and resonant modulators is the crucial need for controlling and tuning their resonance wavelengths~\cite{Dong:10, gevorgyan2021cryo,gehl2017operation,pintus2022ultralow}. This is due to the fact that the absolute resonance wavelength and the resonator's input laser light will not be exactly identical due to process variations. At RT, this challenge can be addressed by thermal tuning of the resonance, using an integrated thermo-optic phase shifter, in a closed- or open-loop fashion~\cite{Dong:10, moazeni201740,Atabaki2018,sun2015single}. However, at cryogenic temperatures (below 4K), the thermo-optic effect is extremely weak (silicon's thermo-optic coefficient degrades from $\sim10^{-4}$ at 300K to $\sim10^{-9}$ at 4K)~\cite{komma2012thermo}. Furthermore, thermo-optical phase shifters require a supply of constant DC currents resulting in large power dissipation, which is undesirable in cryogenic systems due to their limited cooling power budgets. In addition, using a set of separate tunable laser sources~\cite{estrella2021novel} for large-scale systems with hundreds of photonic resonators is not practical. Therefore, it is crucial to develop an alternative tuning method for cryogenic photonic resonators that can provide a sufficiently large resonance tuning range (typically few-$nm$ wavelength range equivalent to over $\pi$ phase-shift) at ultra-low powers, while maintaining the compatibility with today's silicon photonic foundry process for scalability. 


Several methods have been demonstrated to modulate the resonance wavelength of a photonic resonator, including opto-mechanics~\cite{NatPhotonics2022-Piezo}, magneto-optics~\cite{NatElec2022-Magneto}, and electro-optical effects such as the Pockels effect~\cite{eltes2020integrated,han2024cryogenic, youssefi2021cryogenic} and the DC Kerr effect~\cite{chakraborty2020cryogenic}. However, these optical effects are relatively weak cryogenic optical effects that are only suitable for small modulations of the resonance (typically by less than $0.1~nm$ perturbations). Therefore, they either require very large voltages (e.g., more than $50V$ for tuning over even a small range of $1~nm$ resonance shift), or consume enormous electrical power, exhibiting severe pitfalls for cryogenic applications with limited cooling power. Moreover, none of these approaches can be seamlessly integrated with high-speed resonant modulators as an additional resonance tuning knob. Finally, most of these methods require materials that are very challenging to integrate with the foundry silicon photonics. 

In this work, we address the above mentioned challenge by monolithic integration of non-volatile chalcogenide-based phase-change materials (PCM) with silicon photonics to tune a silicon micro-ring modulator (MRM) at sub-4K temperatures. MRMs are one of the most promising resonant photonic modulators for realizing cryogenic optical links and transduction. In addition to enabling ultra-low-power modulation with almost $10~GHz$ bandwidths using the free carrier-plasma dispersion effect~\cite{gevorgyan2021cryo, Gehl:17}, MRMs naturally support wavelength division multiplexing (WDM), allowing simultaneous communication over multiple wavelengths through a single fiber. For cryogenic MRMs, the ring resonance ($\lambda_{MRM}$) needs to be precisely aligned with the laser wavelength ($\lambda_{Laser}$)~\cite{Dong:10, gevorgyan2021cryo,gehl2017operation,pintus2022ultralow}. 
To achieve maximum optical modulation amplitude (OMA), $\lambda_{MRM}$ must be optimally detuned from $\lambda_{Laser}$, a requirement due to the high quality factor (Q-factor) of MRMs (illustrated in Fig.~\ref{fig:fig1}a). This will be done using non-volatile programming of PCM in here.

The PCM has two stable states, amorphous or crystalline, with distinct optical properties and it can be reversibly switched using tailored heat pulses. After switching, the PCM remains in its state in a non-volatile manner, without any static power dissipation. Electrically-programmable PCM have been employed for realizing photonic switches~\cite{fang2021non,rui_review}, directional couplers~\cite{xu2019low}, and optical computing~\cite{Zhou2023, Shastri2021} in RT. Using a prototypical PCM Ge$_{2}$Sb$_{2}$Te$_{5}$ (GST), we demonstrate a large resonance shift of $0.42~nm$ (with minor Q-factor reduction) for an MRM with a free spectral range (FSR) of $4.5~nm$ at 4K temperature. The GST with $12.5~nm$ thickness is deposited on a $8~\mu m$ long section of the MRM. Notably, while the PCM is programmed locally on the sub-$100\mu s$ timescale by heating the device, we demonstrate that the entire chip temperature stabilizes at the base temperature of the cryostat. This work presents the first-ever demonstration of cryogenic non-volatile photonics, where a thin film of GST is electrically switched at sub-4K temperatures. Lastly, we will show a cryogenic optical MRM with closed-loop non-volatile tuning of resonances using GST. We report a modulation bit rate of $+10~Gb/s$ with an extinction ratio (ER) of $4.94~dB$. Our demonstration paves the way for ultra-low power and high performance cryogenic resonant modulators beyond fabrication limitations.

\section{PCM Integrated Micro-ring Modulator}
\label{sec:integration}

\begin{figure}[h]%
\centering
\includegraphics[width=0.95\textwidth]{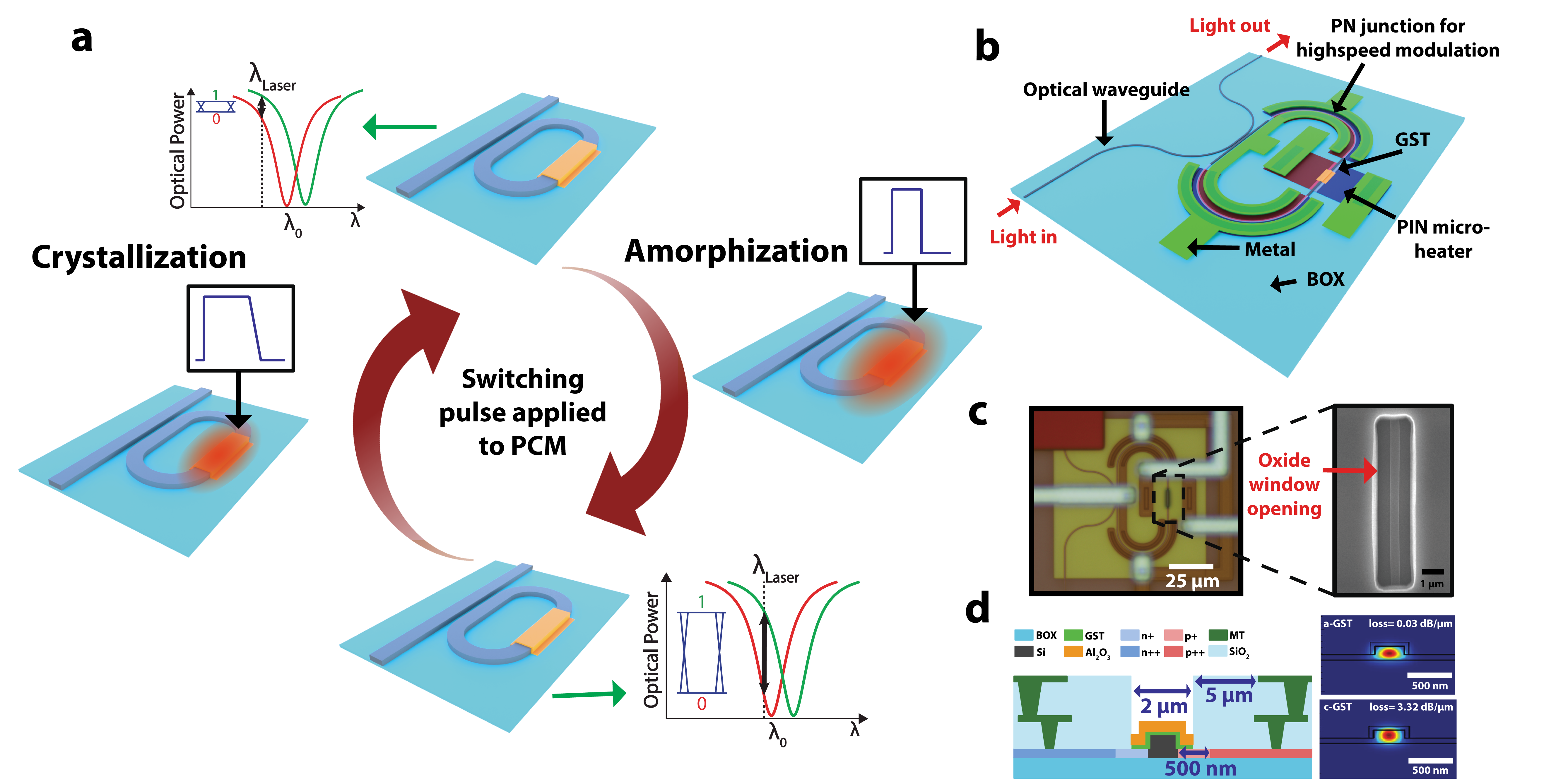}
\caption{(a) The concept of tuning the MRM resonance in a non-volatile fashion using cryogenic PCM, (b) Device illustration of PIN micro-heater integrated with race-track MRM, (c) Cross-section of PIN micro-heater and mode profile simulation with 10nm-thick GST on the waveguide.}
\label{fig:fig1}
\end{figure}

Non-volatile switching in chalcogenide-based PCMs such as GST~\cite{zheng2018gst,zheng2020nonvolatile,rui_gst}, SbS~\cite{delaney2020new,fang2021non,chen2023non}, and SbSe~\cite{rios2021ultra,wu2024freeform,fang2023arbitrary,fang2022ultra} alloys presents a promising solution for post-fabrication resonance tuning of photonic resonators, particularly for cryogenic applications. To achieve this, we propose a PCM-integrated racetrack MRM where the OMA can be maximized by precisely tuning the device to its ``on-resonance" state, that is slightly detuned from the input laser light wavelength as depicted in Fig.~\ref{fig:fig1}a. Electrical programming of the PCM can be performed using a PIN (P-type, Intrinsic, N-type) diode micro-heater structure ~\cite{zheng2020nonvolatile,erickson2022designing,erickson2023comparing}. Our MRM device is designed with a section that includes this embedded PIN micro-heater, while the remaining portion of the resonator enables high-speed modulation using PN junctions based on the free carrier-plasma dispersion effect (Fig.~\ref{fig:fig1}b).

In this work, we utilize the prototypical PCM GST due to its significant optical contrast, with a refractive index of $6.63~+~1.55i$ and $4.6~+~0.34i$ in the crystalline and amorphous states, respectively~\cite{zheng2018gst}. The state of the GST can be reversibly switched by applying a tailored heat pulse that induces a microstructural phase change from the amorphous (a-GST) to crystalline (c-GST) state (the ``SET" process) and vice versa (the ``RESET" process)~\cite{wuttig2017phase}. The GST film can be crystallized by maintaining the temperature above its crystallization temperature ($T_c\sim423$ K~\cite{park1999characterization,morales2002determination,orava2012characterization}) and below its melting temperature ($T_m\sim923$ K~\cite{rui_review}). It can be switched back to the amorphous state when heated above $T_m$ and cooled rapidly. We will study and discuss the impacts of such large temperature requirements during programming of PCM for cryogenic photonics in Section~\ref{subsec:thermal}.

Despite the significant advantages of PCM-based photonics, these materials are not currently available in any commercial silicon photonics. Previous efforts to embed PCM in silicon photonics have required custom modifications to the foundry process~\cite{rios2021ultra,wu2023integration,wei2024monolithic,chen2023deterministic}. Recently, we demonstrated a ``zero-change" post-processing methodology for back-end-of-line (BEOL) monolithic integration of GST into a commercial silicon photonics process~\cite{adya2024post}. We leveraged this approach to realize the proposed MRM devices in a commercial foundry silicon photonics platform using available doping and waveguide features of the process (see Methods).



\section{Measurement Results}
\label{sec:results}

\subsection{Cryogenic MRM Characterization}
\label{subsec2}

\begin{figure}[h]%
\centering
\includegraphics[width=0.95\textwidth]{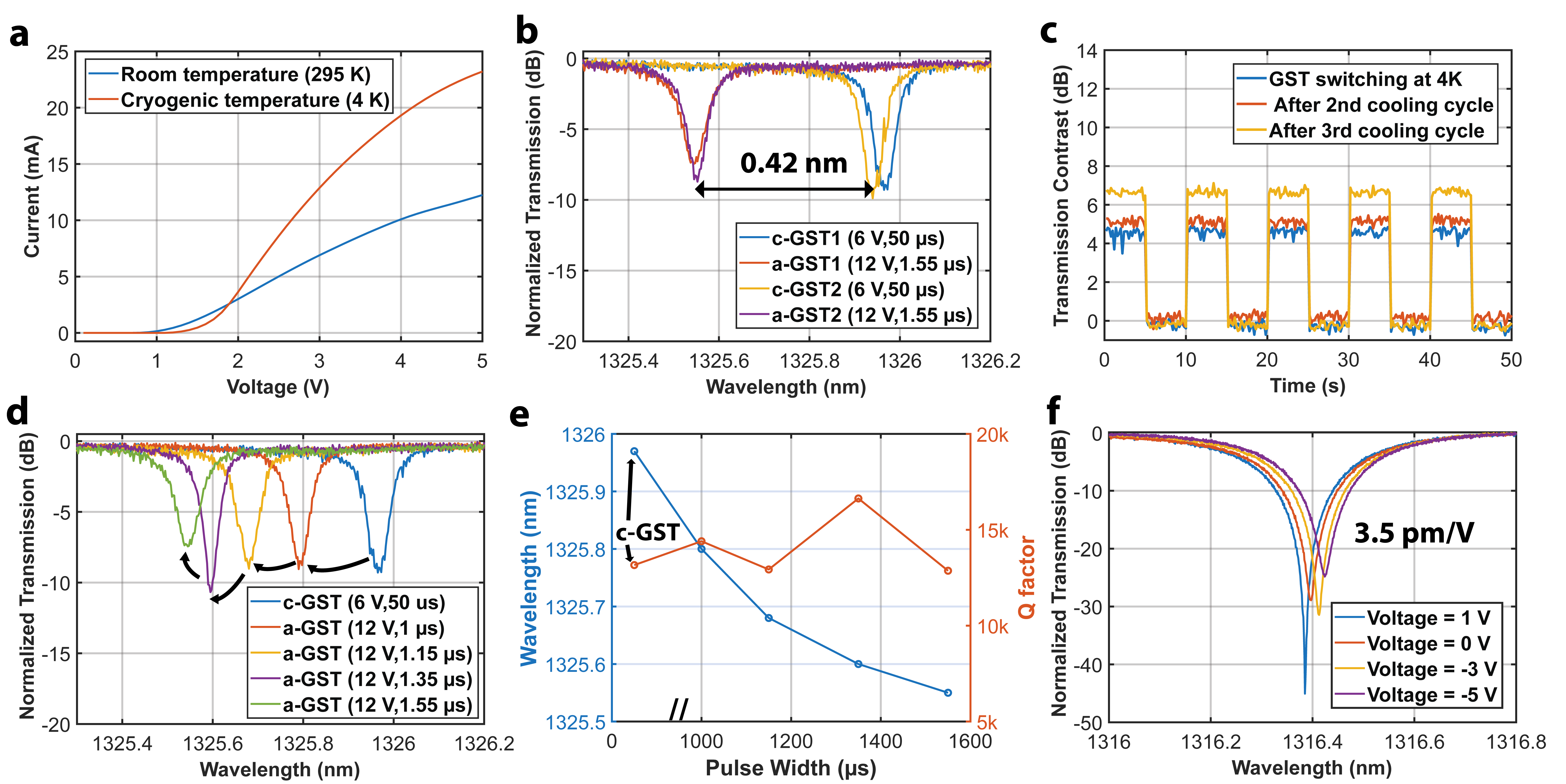}
\caption{PCM tuning characterization: (a) IV characteristics of the PIN micro-heater at RT and 4K, (b) Normalized transmission plot demonstrating the MRM phase shift induced by GST switching at 4K, (c) Change in transmission, near resonance, over consecutive switching events over time and after repeated cooling (from RT to 4K) of the device, (d) Multilevel switching of GST at 4K, (e) Resonance wavelength shift of the MRM during multilevel operation, (f) Optical transmission of the MRM for various voltages applied to the PN junction for high-speed modulation. }
\label{fig:pcm_4k}
\end{figure}

First, we conducted optical measurements to validate the GST switching at cryogenic temperatures and to determine the appropriate switching pulse parameters including pulse voltages, pulse widths, and rise/fall times. We characterized the PIN micro-heater by comparing its I-V characteristics at both RT and 4K, as shown in Fig.~\ref{fig:pcm_4k}a. We observed that the current through the diode was significantly higher, and the threshold voltage was slightly elevated at 4 K as expected due to higher mobility at 4K~\cite{CryoCMOS-IEDM2016, Cryo-electronics}. For characterizing resonance tuning at 4K, we utilized an MRM variant with an $8~\mu m$ oxide window opening, in which a $12.5~nm$ GST thin film was deposited. The MRM device was designed to be in a highly over-coupled state to have relatively low Q-factor for accommodating high electro-optical modulation bandwidths~\cite{moazeni201740}. At 4 K, GST was amorphized by applying a $12V$ pulse for $1.55~\mu s$ (switching energy = $2.09~\mu J$), and crystallized by applying a 6 V pulse for $50~\mu s$ with a $36~\mu s$ falling edge (switching energy = $9.55~\mu J$). Fig.~\ref{fig:pcm_4k}b shows the normalized transmission in the O-band ($1310~nm$) between the two GST states, where we observed a resonance shift of $0.42~nm$ (for an MRM with FSR = $4.5~nm$). The switching is both repeatable and non-volatile, as demonstrated by the binary switching operation in Fig.~\ref{fig:pcm_4k}c. In this experiment, the laser wavelength was tuned near the resonant wavelength for these measurements, and an average ER of $5dB$ was measured. 

We repeated the binary switching measurements each time the device was brought back to RT and then cooled down to 4K. We performed three of such cool-down cycles to assess whether GST switching behavior is affected in any manner. The device consistently operated as a binary switch with only slight variations in ER, indicating that repeated cooling does minimally impact the GST switching properties. However, there is a slight degradation which we believe it is due to the partial switching of GST and can be corrected by appropriately adjusting the switching pulses. Additionally, we performed multilevel switching operations with GST, as depicted in Fig.~\ref{fig:pcm_4k}d. Multilevel switching was achieved by employing pulse width modulation to tune the resonance wavelength from c-GST to a-GST in four steps. The total Q-factor of the MRM is measired to be $\sim~13000$, and it shows minimal variation during multilevel switching, as shown in Fig.~\ref{fig:pcm_4k}e. The minor intrinsic Q-factor reduction from $\sim45k$ in the amorphous state to $\sim40k$ in the crystalline state, suggests an excessive round trip loss of $0.57~dB$. This multilevel switching capability in GST is crucial for fine resonance tuning of the MRM at cryogenic temperatures.

Furthermore, we characterized the free-carrier dispersion effect at 4K (see supplement), and observed a resonance shift efficiency of $3.5~\text{pm/V}$ (Fig.~\ref{fig:pcm_4k}f), similar to the RT measurements. We note that this current modulation efficiency can be improved up to $15~\text{pm/V}$ in this process using highly doped PN junctions~\cite{amf-pam4-100g} which leads to lower Q-factors. More advanced silicon photonics can be utilized in future to achieve $50~\text{pm/V}$ modulation efficiencies~\cite{gevorgyan2021cryo, Rakowski2020}.

\subsection{Thermal Characterization}
\label{subsec:thermal}

\begin{figure}[h]%
\centering
\includegraphics[width=0.95\textwidth]{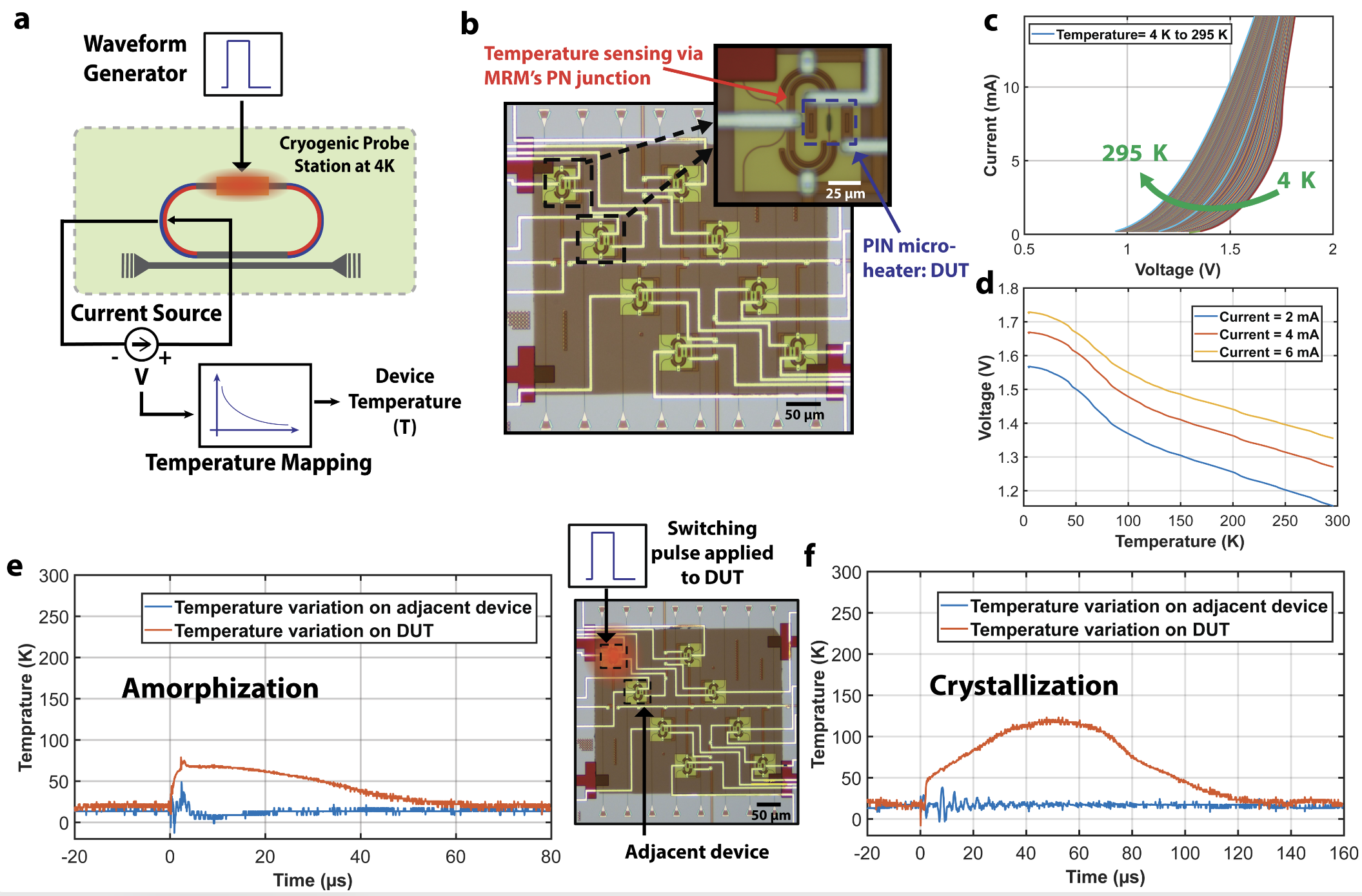}
\caption{Thermal Stress Characterization: (a) Schematic of thermal stress characterization setup, (b) Microscope image of the MRM devices used for thermal stress characterization (PN junctions used as temperature sensors), (c) I-V characteristics of the PN-junctions (temperature sensors) for temperatures from 4K to 295 K, (d) Voltage across the temperature sensors as a function of temperature.  Temperature variations (on device and also on the adjacent device) over time while the amorphization (e) or crystallization (f) pulse is applied to DUT.}

\label{fig:thermal_4k}
\end{figure}

Electro-thermal devices such as thermal phase-shifters are generally not ideal for cryogenic applications as they can exceed the cooling power budget of the cryostat (similarly true for magneto-optical photonics). The heat generated to switch the GST phase is less of a concern to the cooling power of the cryostat since the heat, produced by the short switching pulses, is only generated during the programming pulses and it is localized within a small MRM area. However, one potential concern is heat leakage, which could affect the operation of adjacent thermal-sensitive devices such as superconducting circuits and photodetectors. To characterize the temperature stress around the device during tuning, we utilized the PN junction sections of the MRM as temperature sensors (Fig.~\ref{fig:thermal_4k}b) to record temperature variations when a GST switching pulse was applied to the PIN micro-heater (Fig.~\ref{fig:thermal_4k}a).

The temperature sensor was characterized by performing I-V sweeps at successive temperatures, from 4K to 295K, as shown in Fig.~\ref{fig:thermal_4k}c. Fig.~\ref{fig:thermal_4k}d presents the voltage recorded across the temperature sensor as a function of temperature. After applying the pulse to the PIN micro-heater, the temperature was measured at two locations: on the same MRM device under test (DUT) as the PIN micro-heater is being programmed, and on an adjacent MRM device located $50~\mu m$ away. When the PCM switching pulse was applied, a negligible temperature rise was detected on the adjacent device that was located $50~\mu m$ away, as indicated by the blue plot in Fig.~\ref{fig:thermal_4k}e and f. 

In contrast, a temperature rise was observed when the PIN micro-heater and the temperature sensor were co-located at the same ring (Fig.~\ref{fig:thermal_4k}a). During crystallization, the temperature rose to 120K and dropped back to 4K within $140~\mu s$ (Fig.~\ref{fig:thermal_4k}f), while during amorphization, the temperature rose to 75K and returned to 4K within $60~\mu s$ (Fig.~\ref{fig:thermal_4k}e). The temperature measured here will correspond to the average temperature across the ring while only the PIN micro-heater section of the MRM is switched, suggesting that the temperature rise is highly localized within the same device returning to to 4K within a few microseconds. Moreover, the temperature rise has a negligible effect beyond $50~\mu m$ from the PIN micro-heater.

\subsection{Cryogenic Optical Modulator with Resonance Tuning}



\begin{figure}%
\centering
\includegraphics[width=0.95\textwidth]{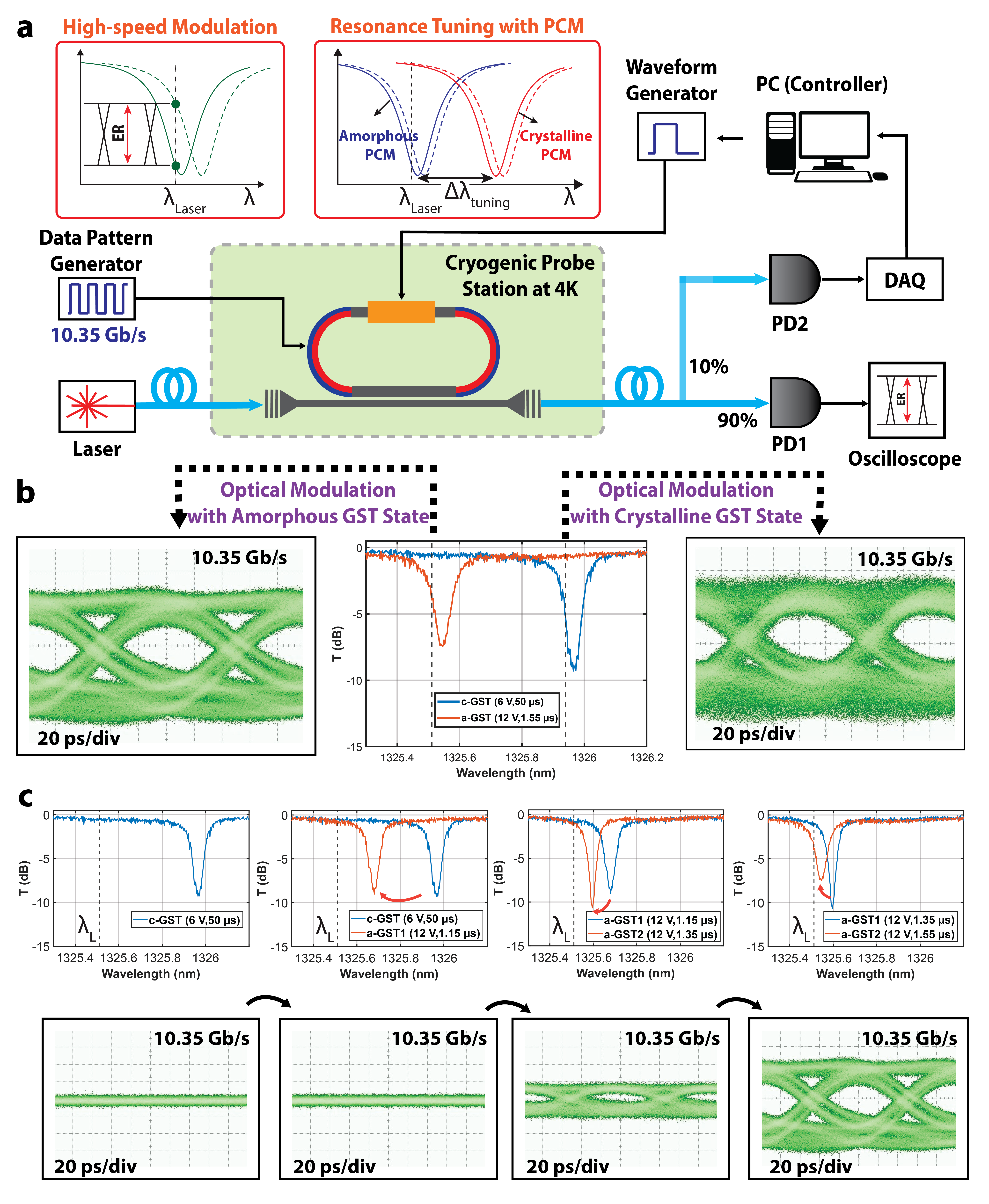}
\caption{Demonstration of resonance tuning in cryogenic MRM using GST: (a) Block-diagram of the closed-loop resonance tuning and locking mechanism using GST programming, (b) High-speed optical data transmission measurements at a-GST on-resonance point and c-GST on-resonance point,(c) Gradual tuning and locking of MRM to its on-resonance state using multilevel GST programming.}
\label{fig:system_4k}
\end{figure}

Lastly, we demonstrate an optical transmitter operating at cryogenic temperatures that included the racetrack MRM for high-speed modulation, and a PCM-based closed-loop resonance tuning scheme. In this experiment, we demonstrate how non-volatile reconfiguration of the MRM's resonance via GST enables successful data modulation with maximum OMA and ER, despite initial misalignment between the laser and MRM resonance wavelengths.
To characterize the modulation performance of the PCM-integrated MRM, we performed optical high-speed data transmission measurement in a 4K cryogenic electro-optical probe station. A high-speed electrical signal was applied to the MRM's PN junctions at a data rate of $10.35~Gb/s$ and applied voltage levels of $0V$ and $-3.6V$, ensuring depletion-mode operation for high-speed data modulation~\cite{gevorgyan2021cryo} using a random bit pattern generator. We characterized the high-speed modulation performance of the MRM in either states of PCM. When the modulator was operated at the on-resonance point with GST in its amorphous state ($\lambda_{MRM} = 1325.511nm$), an eye diagram was captured using an external high-speed photodetector (PD), as shown in Fig.~\ref{fig:system_4k}b. We measured an ER of $4.94dB$ and an insertion loss (IL) of $2.73~dB$. Next, the GST was switched to its crystalline state, and the laser was tuned to the new on-resonance point ($\lambda_{MRM} = 1325.939nm$), yielding an ER of $3.81~dB$ and an IL of $3~dB$. This shows we can achieve a high data rate modulation regardless of the PCM state.

To demonstrate the non-volatile tuning and locking of the MRM resonance using GST, the laser wavelength was fixed at an arbitrary wavelength, around $\sim1325.5nm$ in this case near the amorphous on-resonance point. The GST was initially set to its fully crystalline state, where the laser wavelength is far off-resonance, and no eye-opening was observed on the oscilloscope. The modulator was then tuned using partial-amorphization pulses to perform multilevel switching to eventually bring the modulator closer to the on-resonance point with a closed-loop monitoring setup (see Methods). The tuning loop works based on monitoring the average optical power from the output of MRM (via a second PD and a 10/90 coupler), and locking the average optical power at the output of modulator to an ideal pre-calculated values based on initial characterizations of the MRM with a target locking range.

After the first partial-amorphization pulse (12 V, $1.15~\mu s$), no eye-opening was observed as there was still a large mismatch between the laser wavelength and resonance, so a second pulse was applied. After the second pulse (12 V, $1.35~\mu s$), a partial eye-opening was detected, prompting the application of a final pulse (12 V, $1.55~\mu s$). With this final amorphization pulse, the maximum eye-opening was achieved by locking the modulator in its on-resonance state. 

Tuning of the resonant modulator to its on-resonance state can be achieved using the non-volatile nature of GST where it self-holds the state with zero static power once set to a particular state. By starting from the crystalline state, corresponding to the highest $\lambda_{MRM}$, and sweeping the resonance to lower wavelengths we ensure that the $\lambda_{MRM}$ will be always larger than $\lambda_{Laser}$. This is critical to avoid any self-heating instability in MRMs~\cite{moazeni201740}. This tuning procedure can be repeated for resonance alignment to any arbitrary laser wavelength within the GST tuning range, with step sizes adjusted according to the resonator’s Q-factor and accuracy of locking point.

\section{Discussion}
\label{sec:conclusion}

We have demonstrated non-volatile tuning of cryogenic resonant modulators using PCM. This approach provides an ideal tuning mechanism to align hundreds of resonators to desired wavelengths with zero static power dissipation, addressing both process variations and thermal drifts. As a result, it enables dense WDM (DWDM) optical I/O with aggregate bandwidths in the Tb/s range in a single fiber, supporting ultra-low power and high-speed optical links between 4K and RT environments for cryogenic classical and quantum computing.

The first-ever non-volatile cryogenic photonic device presented in this work exhibits electrically programmable tuning with high repeatability (see Supplementary Information). We observed that the pulse widths required for GST amorphization are approximately $10\times$ longer at 4K compared to room temperature, while similar voltage levels are maintained. This behavior is likely due to the cryogenic environment, which facilitates rapid cooling of the melted GST and prevents re-crystallization. Longer amorphization times are advantageous, offering more precise PCM switching to desired states and allowing control over thicker PCM films (see Supplementary Information).

Additionally, we characterized the temperature stress around the PIN micro-heater during PCM programming and found no significant temperature rise beyond a $50~\mu$m radius. This ensures that temperature-sensitive devices, such as superconducting nanowire single-photon detectors, can be integrated in close proximity to the proposed resonators without interference. We also demonstrated resonance tuning and locking in a PCM-integrated racetrack MRM, achieving an optical data transmission rate of $10~\text{Gb/s}$. Although optimizing the electro-optical bandwidths of modulation at 4K was beyond the scope of this work, this can be improved using higher doping concentrations and optimized junction geometries.

The device was gradually tuned to its on-resonance state through multilevel GST programming at cryogenic temperatures. The current device achieves a tuning range of $0.42~\text{nm}$, limited by the thickness and length of the GST layer. However, this range could be extended to over $7~\text{nm}$ with minimal impact on the Q-factor by using low-loss PCM such as SbS and SbSe at longer length sections (see Supplementary Information). This range is sufficient to cover the FSR of typical MRMs in majority of applications.

Moreover, wide-bandgap, low-loss PCM-based switches can control single-photon paths at cryogenic temperatures, opening new possibilities for quantum photonics. While this study focuses on tuning MRMs using cryogenic PCMs, the same method can be extended to cryogenic optical receivers for WDM operation. The non-volatile photonics enabled by PCM in this work paves the way for cryogenic integrated photonics, featuring thousands of resonant devices operating at ultra-low power with zero static power dissipation, facilitating future applications in quantum technologies, optical and ultra-low noise computing, and high-energy physics.

\section*{Methods}
\label{sec:methods}

\textbf{Silicon Photonic Chip Fabrication.} We utilized a multi-project wafer (MPW) program to design our silicon photonic chips using the Advanced Micro Foundry (AMF) commercial foundry process. A rib waveguide was implemented in our PIN-micro-heater design to enable single-mode operation in the O-band ($1310~nm$). We leveraged an existing feature of this technology to open an oxide window down to the silicon layer. While typically used for bio-sensing applications, this method allowed us to deposit the PCM on the desired waveguide sections. A racetrack MRM was designed with a PIN-micro-heater incorporating a $2~\mu m$-wide oxide window opening. Since the oxide etch process at the foundry could potentially damage surrounding metal routings, we ensured that all non-silicon layers, including metal and via layers, were positioned at least $5~\mu m$ away from the oxide opening area. This arrangement resulted in a high overall resistance of the micro-heater's parasitic junction.

To achieve low optical losses in the waveguide, we introduced low doping (n+/p+) in the slab sections near the waveguide core, while increasing the doping concentration further from the core to reduce overall device resistance and maintain low contact resistance at the vias. After receiving the dies from the foundry, each die was post-processed using coarse-resolution lithography for the monolithic integration of GST (see Supplementary Information). The bent regions of the racetrack MRM have a radius of $12.5~\mu m$; as shown in Fig.~\ref{fig:S4}a, the waveguide core was doped with the lowest (n/p) concentration, with the doping level increasing further from the core. The coupler between the MRM and the bus waveguide featured a coupling gap of $350~nm$ and a coupling length of $2~\mu m$. Based on FDTD simulations, this combination of gap and length was estimated to couple approximately 13.75 percent of the power from the bus waveguide into the ring (Fig.~\ref{fig:S4}b).

\textbf{Device Characterization and Testing.} We conducted the PCM tuning characterization using an electro-optical cryogenic probe station (Lakeshore CRX-4K). Light was coupled to the resonators via grating couplers, resulting in a loss of 6 dB per coupler at 4K. Single DC probes were used to contact the electrical terminals of the PIN micro-heater. The test setup included an O-band tunable laser source (Santec TSL-550), power meter (Agilent 81635A), photo-detector (Thorlabs PDB450C), DAQ, source meter (Keithley 2604), and waveform generator (Keysight 33522B). The waveform generator produced the PCM switching pulses. Thermal characterizations were performed by wire-bonding the electrical contacts of the MRM PN junctions to a PCB, using them as temperature sensors. We electrically probed the PIN micro-heater while biasing the temperature sensors (MRM PN junction) with a current source, and recorded voltage variations across them with an oscilloscope. 

For high-speed optical data transmission measurements, a bit pattern generator (Keysight N4903A Serial BERT) provided the electrical data input, and the optical output was monitored on an oscilloscope (Keysight 86100D DCA-X) through a high-speed photo-detector (Thorlabs RXM10AF). The high-speed electrical signal was sent to the modulator via RF probes, while single DC probes were used for the PIN micro-heater (Fig.~\ref{fig:S5}).  We measured the ER and IL by directly observing the optical power on an optical scope. An additional PD was employed to monitor and sense the optical power inside the resonator. The PD output was digitized and read by a data acquisition (DAQ) unit, which sent the data to a PC controller. The controller processed the data to continuously tune the GST state and adjust $\lambda_{\text{MRM}}$, bringing it close to $\lambda_{Laser}$. This tuning approach ensures that $\lambda_{Laser} < \lambda_{MRM}$, avoiding potential self-heating stability issues~\cite{moazeni201740}.

\section*{Acknowledgments}

We would like to acknowledge CMC Microsystems for providing MPW fabrication services for AMF Silicon Photonics technology. Post-processing steps were conducted at the Washington Nanofabrication Facility (WNF), a National Nanotechnology Coordinated Infrastructure (NNCI) site at the University of Washington, with partial support from the National Science Foundation via awards NNCI-1542101 and NNCI-2025489.

This work is supported by National Science Foundation (NSF) under Grant No. CCF-2105972, the U.S. Department of Energy (DOE) under Award Number DE-SC0024729, and ONR MURI (award no. N00014-17-1-2661). A.M and R.C. are supported by NSF-FuSe Award Number 2329089.

\bibliography{main}

\newpage
\section{Supplementary Information}
\label{sec:supplementary}

\subsection{PCM Post-processing} 
\begin{figure}[h]%
\centering
\includegraphics[width=1\textwidth]{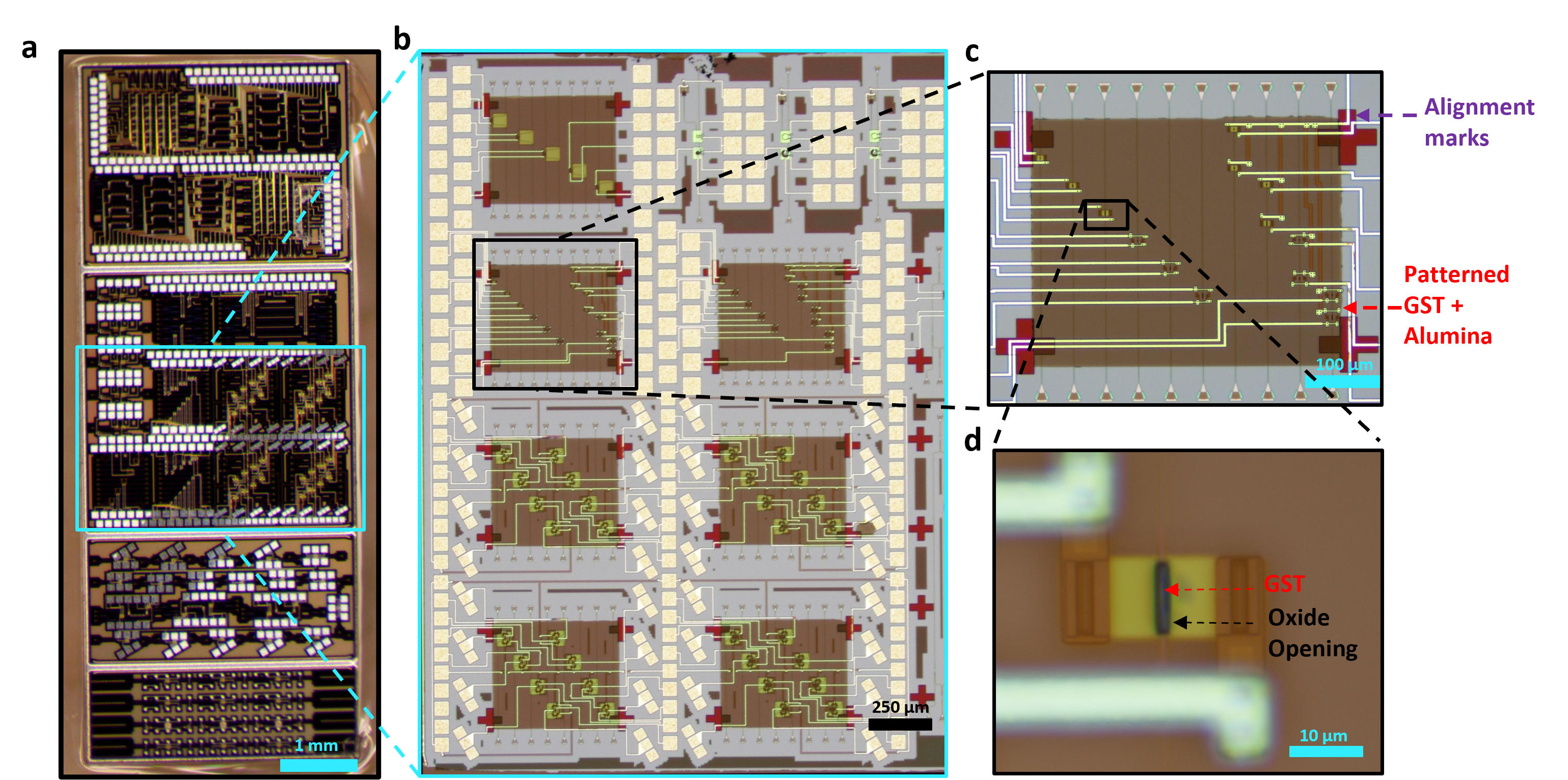}
\caption{(a) Microscope image of the unprocessed chip, (b) microscope image of post-processed section of the chip, (c) microscope image of a single post-processed section, and (d) the PCM micro-heater with oxide opening for GST deposition.}
\label{fig:S1}
\end{figure}

We monolithically integrated GST with the AMF silicon photonics chip using a few post-processing steps at the die level ($3~mm \times 3~mm$) in the Washington Nanofabrication Facility (WNF). This post-processing method is easily scalable, as it involves a minimal number of steps and uses coarse-resolution lithography. We began by mounting the die onto a large Si carrier chip for ease of handling, using AZ10XT, and then spin-coated it with AZ1512 photoresist. The die was patterned using the Heidelberg DWL66+ laser writer to block metal pads and grating couplers during GST deposition. After developing the patterned chip in AZ MIF developer, we sputtered a $12.5~nm$ thick GST film, followed by a $60~nm$ thick $Al_2O_3$ layer using atomic layer deposition (ALD) at $323$ K, to prevent crystallization of GST, which occurs around $423$ K. Lift-off was performed as the final step to remove residual photoresist. Fig.~\ref{fig:S1}a shows the complete chip layout before post-processing. The blanket deposition of GST makes this technique suitable for integration with large-scale programmable circuits. In our layout design, multiple device batches were placed in a grid pattern, as shown in Fig.~\ref{fig:S1}b, which displays only the post-processed section of the chip. Fig.~\ref{fig:S1}d provides a microscope image of a single PIN micro-heater device with an $8~\mu m$-long oxide window opening.


\subsection{Endurance Test of GST at Cryogenic Temperatures}

We conducted GST endurance measurements to evaluate its lifespan at cryogenic temperatures. For these measurements, we used an MRM variant with a $4~\mu m$ oxide window opening and a $12.5~nm$ GST thin film deposited in the opening. A resonance shift of 0.15 nm was observed when switching the GST states. To perform the endurance test, we first tuned the laser wavelength close to the device’s resonance wavelength with the GST in its amorphous state. Next, we recorded the transmission contrast at that laser wavelength across several GST switching events. Although each switching event occurs as quickly as $\sim 100~\mu s$, we introduced a $5~s$ delay between events to account for instrument delays and to ensure the device had sufficient time to cool to 4K.

\begin{figure}[h]%

\centering
\includegraphics[width=0.95\textwidth]{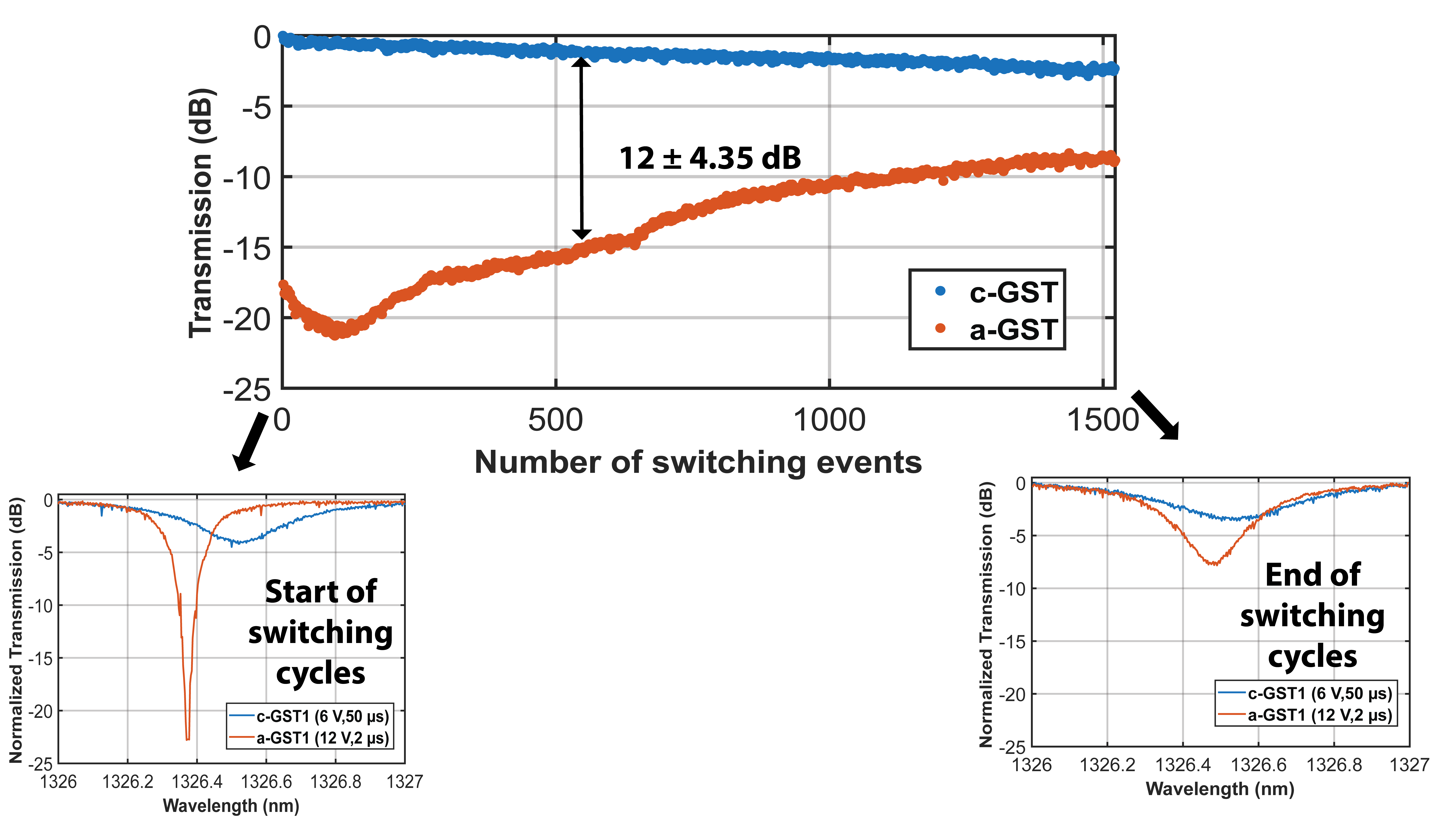}
\caption{The endurance measurement results of GST-integrated MRM at 4K for 1500 switching events with normalized recorded transmission shown at the start and end of the measurement.}
\label{fig:life_4k}
\end{figure}

We recorded 1500 switching events and observed a 10 dB variation in the a-GST transmission (Fig.~\ref{fig:life_4k}). Analysis of the transmission plots showed that, after 1500 switching events, the extent of switching behavior diminished, with a reduction in Q-factor indicating additional loss in the device. We attribute these losses to potential GST degradation and reflow issues. Furthermore, it is likely that, over time, different pulse conditions may be necessary to achieve consistent switching; maintaining constant pulse conditions throughout the endurance test may have led to inadequate switching behavior. Future studies can provide a deeper understanding of GST endurance by further characterizing these phenomena and de-embedding the various loss mechanisms contributing to this behavior. GST lifespan may be enhanced by using patterned GST deposition only on the waveguide, followed by full encapsulation with an alumina layer to prevent any melted GST from re-flowing.

\subsection{Switching Thick-GST Films at Cryogenic Temperatures}

\begin{figure}[h]%
\centering
\includegraphics[width=0.95\textwidth]{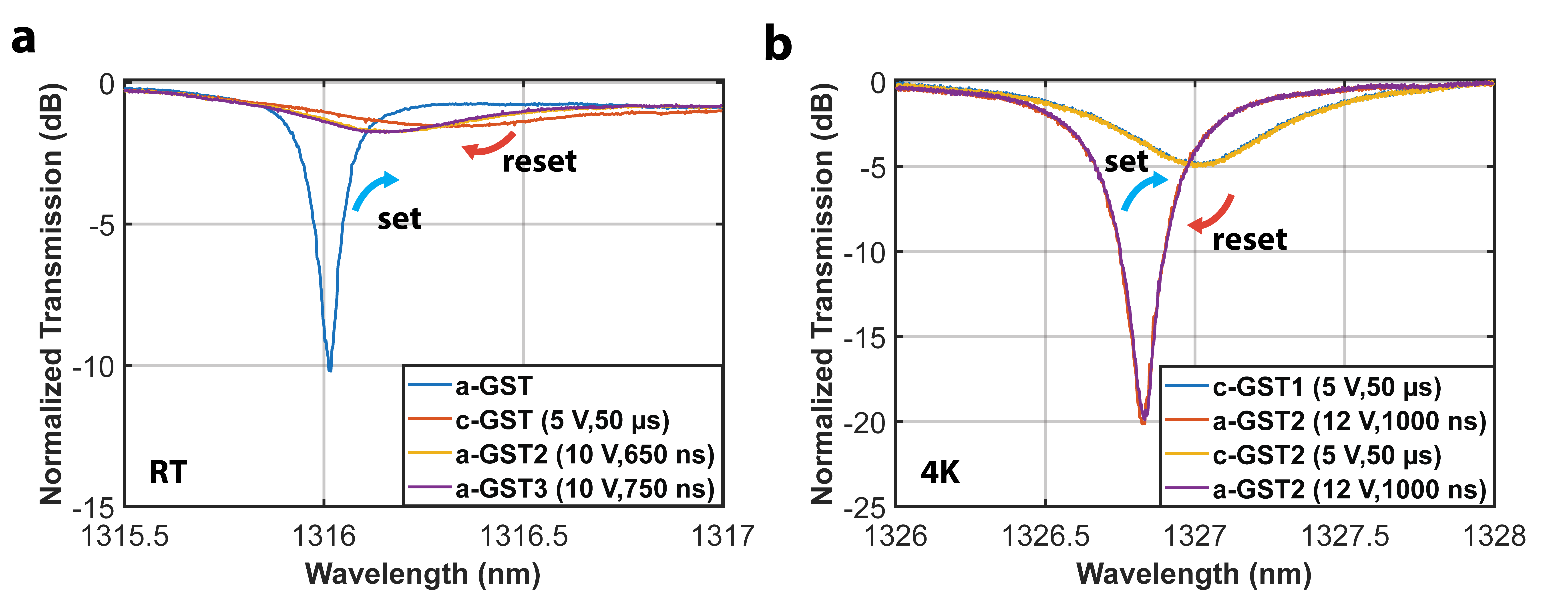}
\caption{(a) Normalized transmission plot demonstrating one-time switching of 20 nm thick GST at RT, (b) Normalized transmission plot demonstrating two switching cycles of 20 nm thick GST at 4K.}
\label{fig:S2}
\end{figure}

We conducted GST tuning measurements of thick GST films at cryogenic temperatures. As previously noted, the cryogenic environment facilitates rapid cooling of melted GST and prevents re-crystallization during the amorphization process, allowing for much longer pulse durations compared to RT. For these measurements, we used an MRM variant with a $4~\mu m$ oxide window opening and a $20~nm$ GST film deposited in the opening. During RT measurements, we did not observe repeatable switching and found it challenging to re-amorphize the $20~nm$ thick GST film, even with very long pulse durations, as shown in Fig.~\ref{fig:S2}a. This behavior may be attributed to insufficient and non-uniform heating of the entire GST volume over short amorphization timescales, potentially leading to switching issues or material degradation. 

We repeated the switching measurement at 4K on a different device. At 4K, we observed precise and repeatable switching behavior (Fig.~\ref{fig:S2}b) in the $20~nm$ thick GST film, with a resonance shift of $0.2~nm$. Here, GST was amorphized by applying a $12~V$ pulse for $1~\mu s$ and crystallized by applying a $5~V$ pulse for $50~\mu s$. This confirms that longer amorphization pulse durations in a cryogenic environment can provide effective control over thicker PCM films, enabling precise and repeatable switching operations.


\subsection{Comparison of Cryogenic Photonic Mirco-ring Modulators}
We have compared various resonator-based cryogenic photonic modulators in Table.~\ref{tab1}. There are several recent modulation methods demonstrated at cryogenic temperatures operating at as high as $20~Gb/s$ but so far none of these modulators have had an independent phase-shifting capability for tuning the resonance wavelength. We discuss why none of the previous approaches are suitable for tuning the resonators below:

Magneto-optical effects have been reported recently~\cite{NatElec2022-Magneto} operating at $2~Gb/s$, but they require large electrical currents in the order of $100 mA$ range for modulation of resonance with less than $50pm$ shift. Hence, while this effect is stronger than thermo-optics at 4K (by about $5\times$), it still cannot provide large wavelength ranges for resonance tuning. Additionally, the integration of required materials for magneto-optics with foundry silicon photonics is extremely challenging. 

The Pockels effect in BaTiO$_{3}$~\cite{eltes2020integrated} has also been used in cryogenic devices for modulation at $20~Gb/s$, but tuning a device using this method over sufficiently large ranges requires applying extremely large voltages of $\sim100V$ (resonance shift of $0.645~nm$ would require $150V$). Similarly, there have been other demonstrations of high-speed energy-efficient cryogenic photonic modulators in Silicon including micro-ring modulators (MRM)~\cite{gevorgyan2021cryo}, micro-disk modulators~\cite{gehl2017operation}, and InP-on-Si quantum well MRM~\cite{pintus2022ultralow}, but yet these lack a strong enough effect for resonance tuning. In the Si-spoked MRM~\cite{gevorgyan2021cryo}, its free-carrier plasma-based tuning range of $0.8~nm$ can be potentially utilized for resonance tuning across small wavelengths. However, this still lacks having two independent control knobs for high-speed operation as well as resonance tuning, which is a limiting factor. 

Additionally, there have been notable demonstrations of non-resonator-based modulators with footprints of $10\times$ to $100\times$ larger than resonant modulators. These devices are either electro-absorption based~\cite{chansky2022high}, or Mach-Zehnder modulators (MZM) based on Pockels effect using LiNbO$_{3}$~\cite{youssefi2021cryogenic,han2024cryogenic}, opto-mechanics using piezo-electric effect in AlN~\cite{NatPhotonics2022-Piezo}, DC Kerr effect~\cite{chakraborty2020cryogenic}, and silicon-organic hybrid (SOH) MZM~\cite{schwarzenberger2022cryogenic}. While these devices could achieve data rates as high as $50~Gb/s$, their area and energy inefficiencies will be limiting their use in future cryogenic computing systems.


\subsection{Discussion of Tuning Range using Ultra-low Loss PCM} 
In practical applications, the resonance of an MRM should be tunable across the required wavelength range, which varies depending on the optical link system's use case~\cite{Dong:10, moazeni201740,Atabaki2018,sun2015single}. For instance, to account for fabrication variations in absolute laser wavelengths or resonator shifts, a tuning range of $1~nm$ to $2~nm$ may be sufficient. In dense WDM (DWDM) links, this range can adequately cover the channel-to-channel spacing (typically $1~nm$ to $2~nm$). However, ideally, the entire FSR range should be tunable to ensure robust modulation regardless of temperature shifts, process variations, and other artifacts. While our device currently achieves a tuning range of $0.42~nm$ (for a resonator FSR = $4.5~nm$), this can be easily increased by using a longer GST length or increasing its thickness to induce a larger phase shift. A significant drawback of using GST for resonance tuning, however, is its high optical loss in the crystalline state, which limits the achievable phase shift without a substantial drop in the cavity's Q-factor. To address this, ultra-low loss PCMs, such as SbSe or SbS, could be used to achieve phase shifts and significantly extend the tuning range.

In this work, we simulated a $\sim~20~nm$ thick, low-loss PCM to estimate the phase shifter length required for a full FSR shift. For a ring with a radius of $12.5~\mu m$, a full FSR shift ($\sim~7.3~nm$) was achieved with a $16~\mu m$ long SbSe-based phase shifter, showing minimal changes in Q-factor and ER between the amorphous and crystalline states (Fig.~\ref{fig:S4}a). Similarly, we see that a half FSR shift was achieved with a $8~\mu m$ long SbSe-based phase shifter. For an SbS-based phase shifter, a longer length of $36~\mu m$ is required to achieve a full FSR shift, due to a smaller effective refractive index contrast, and a $18~\mu m$ phase-shifter achieves half FSR shift. Higher absorption loss in SbS resulted in a more significant variation in Q-factor and ER (Fig.~\ref{fig:S4}b). Overall, these simulation results indicate a promising approach to extend the tuning range up to ($\sim~7.3~nm$) in the future by using low-loss PCM, increasing the PCM section length, and optimizing PCM thickness.

\newpage

\newpage
\section*{Figures}
\label{sec:figures}

\newpage

\newpage
\clearpage
\section{Supplementary Data}
\label{sec:supplementary_data}

\begin{table}[h!]
\centering
\label{tab1}
\footnotesize
\resizebox{1\textwidth}{!}{
\begin{tabular}{ p{0.1\linewidth} p{0.2\linewidth} p{0.1\linewidth} p{0.1\linewidth} p{0.1\linewidth} p{0.1\linewidth} p{0.17\linewidth}}
\hline

 \textbf{Reference} & \textbf{Modulation Method} & \textbf{Device} & \textbf{Data rate} & \textbf{Modulation} & \textbf{Electrical Energy} & \textbf{Resonance} \\ 
  \textbf{} & \textbf{} & \textbf{Structure} & \textbf{(Gb/s)} & \textbf{Voltage (V)} & \textbf{(fJ/bit)} & \textbf{Tuning} \\ 
\hline
\cite{NatElec2022-Magneto}     & Magneto-optic effect  & Micro-ring  & 2 & -  & 3900 & NA \\ 

\cite{eltes2020integrated}     & Pockels effect in BaTiO$_{3}$  & Micro-ring  & 20 & 1.7  & 45 & NA \\ 

\cite{lee2020high}     & Graphene on SiN   & Micro-ring  & 20 & 3  & 20 & NA \\ 

\cite{pintus2022ultralow}     & InP-on-Si QW    & Micro-ring  & 4 & 0.11  & 1.3 & NA \\ 

\cite{gehl2017operation}    & Free-carrier plasma dispersion in Si   & Micro-disk  & 10 & 1.8  & 80 & NA \\ 

\cite{gevorgyan2021cryo}     & Free-carrier plasma dispersion in Si    & Micro-ring  & 20 & 1.5  & 54 & NA$^*$ \\ 

\textbf{This work}    & Free-carrier plasma dispersion in Si    & Micro-ring  & 10 & 3.6  & 75 & Non-volatile using PCM \\ 
\hline
\end{tabular}}

\caption{Comparison of various cryogenic resonant modulators (NA: Not Applicable, *Resonance cannot be independently tuned in this device).}
\end{table}

\begin{figure}[h]%

\centering
\includegraphics[width=0.75\textwidth]{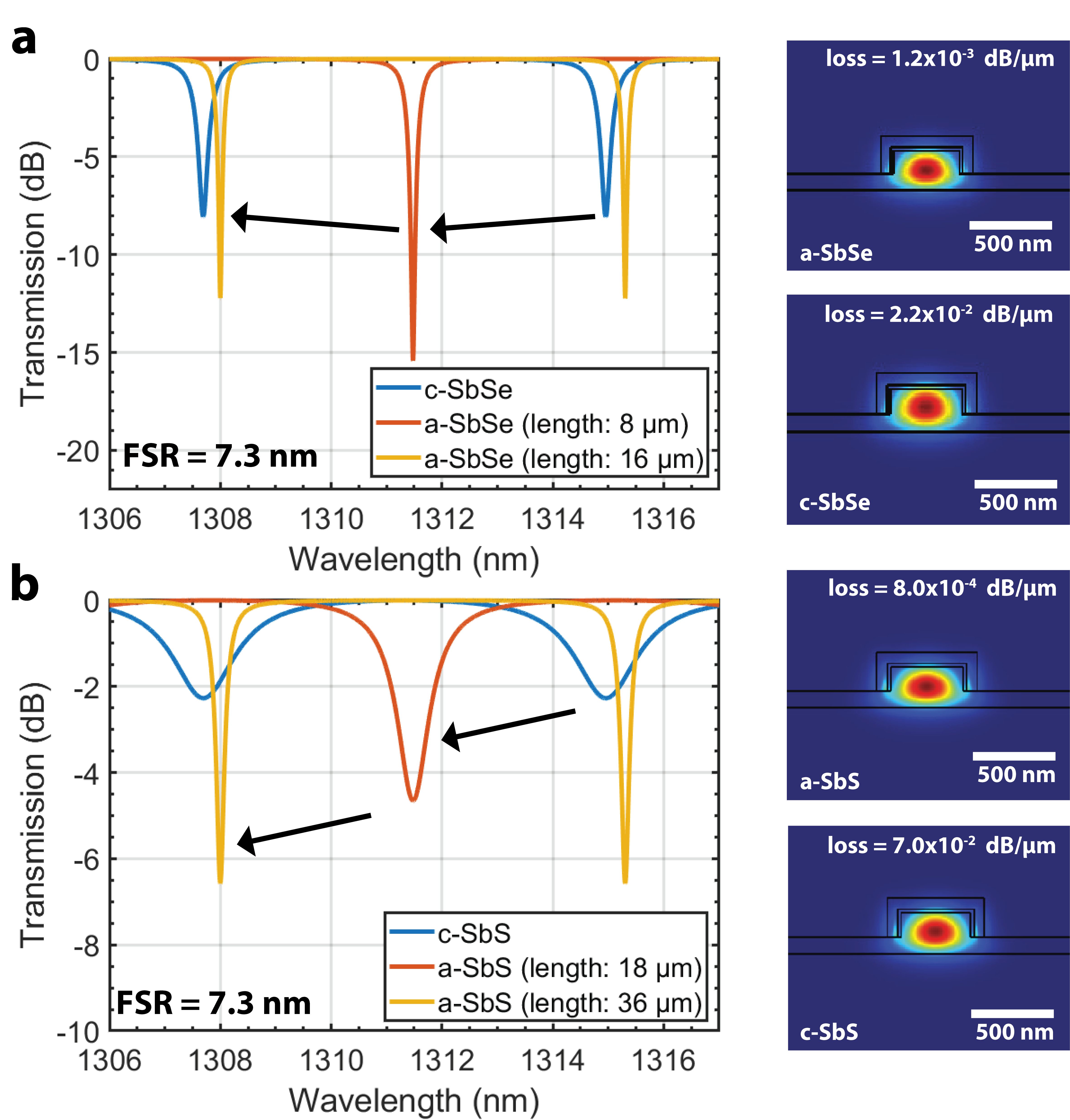}
\caption{Transmission plots and mode profile comparing a full FSR shift induced using (a) SbSe and (b) SbS}
\label{fig:S3}
\end{figure}

\begin{figure}[h]%
\centering
\includegraphics[width=0.75\textwidth]{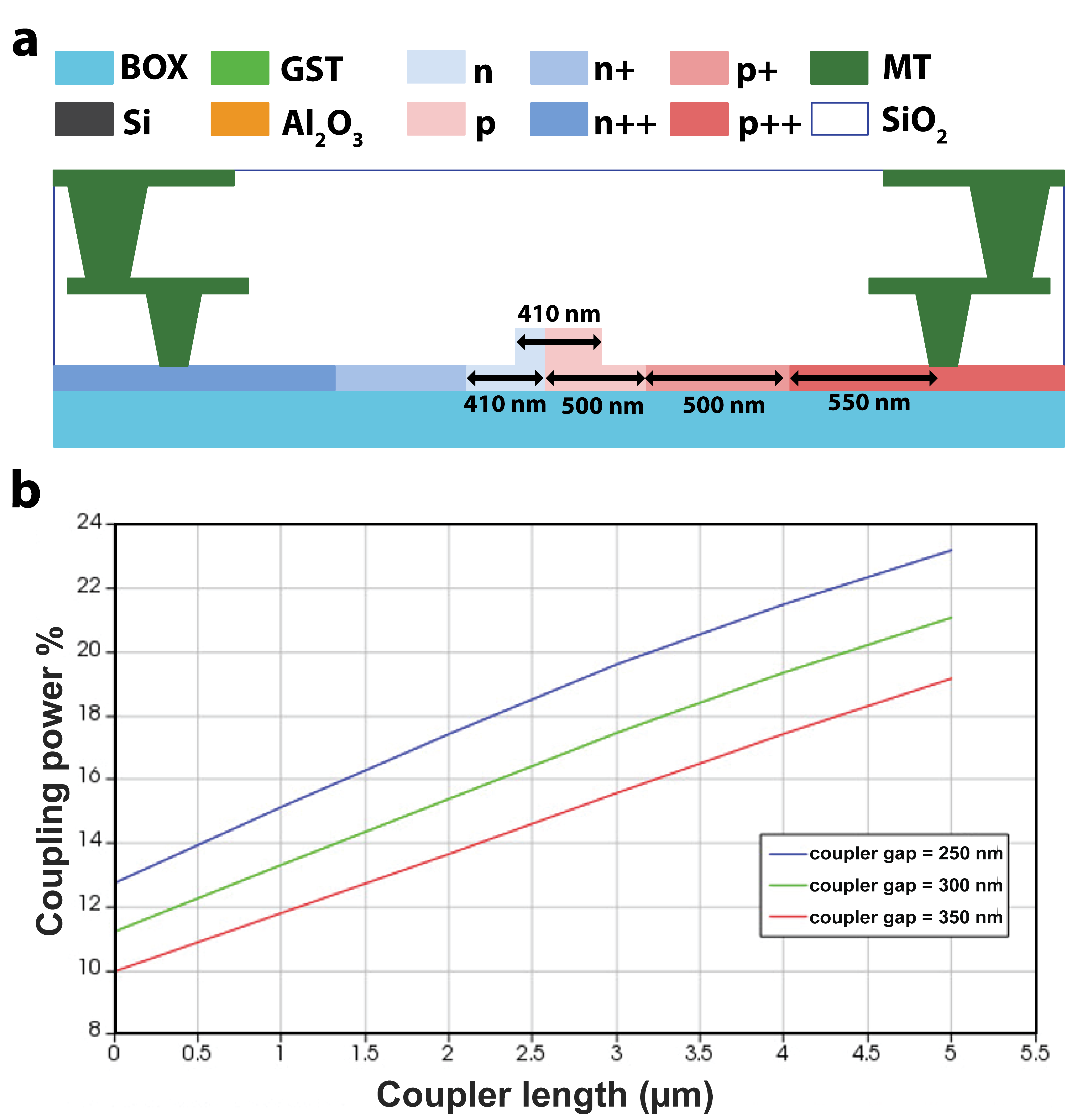}
\caption{(a) Cross-section of the MRM PN junctions, (b) Simulated coupling power percent as a function of coupler length.}
\label{fig:S4}
\end{figure}

\begin{figure}[h]%
\centering
\includegraphics[width=1\textwidth]{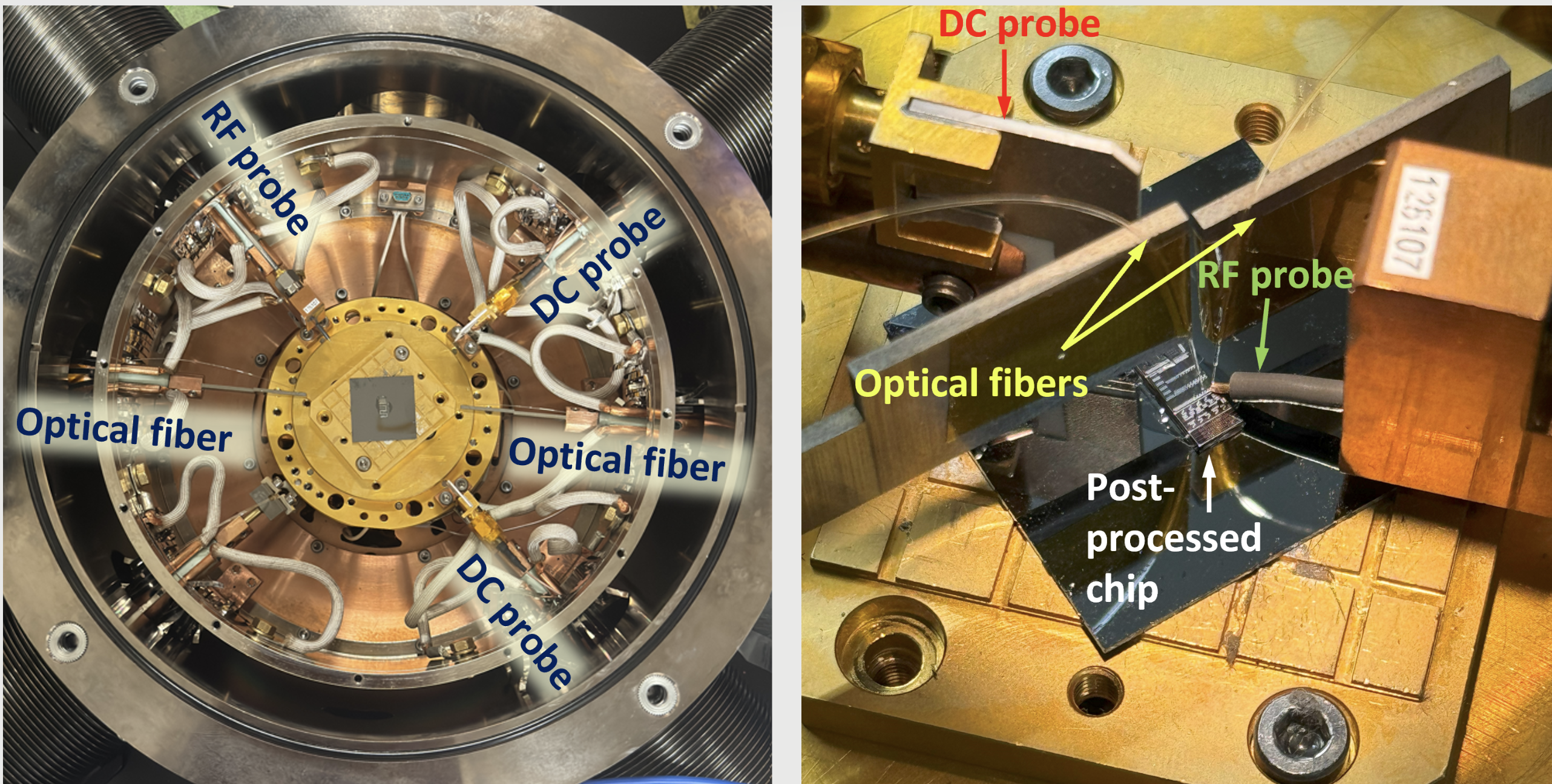}
\caption{Cryogenic measurement test setup showing the post-processed chip, optical fibers, DC and RF probes.}
\label{fig:S5}
\end{figure}

\end{document}